\documentclass[11pt, draftclsnofoot, onecolumn]{IEEEtran}
\usepackage[utf8]{inputenc}
\usepackage{graphicx} 
\usepackage{dblfloatfix}
\usepackage{float}
\usepackage{tabularx}
\usepackage{placeins}
\usepackage{color}
\usepackage{array}
\usepackage{colortbl}
\usepackage{multicol}
\usepackage{lipsum}
\usepackage{graphicx,cite}
\usepackage{amssymb}
\usepackage{array}
\usepackage{amsfonts}
\usepackage{longtable}
\usepackage[aboveskip=.5cm]{caption}

\usepackage{xcolor,colortbl}
\usepackage{booktabs} 
\usepackage{pdfpages}
\usepackage{hyperref} 
\hypersetup{	
colorlinks=true, 
breaklinks=true, 
urlcolor= blue, 
citecolor=black,	
pdftitle={survey}, 
pdfauthor={Anonyme}, 
pdfsubject={Simulation}	
}

\begin{document}

\title{Multi-Criteria Virtual Machine Placement in Cloud Computing Environments: A literature Review}

\author{\IEEEauthorblockN{Wissal Attaoui$^{\dag}$, \it{Student Member, IEEE}, Essaid Sabir$^\dag$, \it{Senior Member, IEEE}\\}
\IEEEauthorblockA{
$^\dag$Hassan II University, Casablanca, Morocco\\
Emails: wissal.attaoui@gmail.com, e.sabir@ensem.ac.ma}}
\maketitle


\begin{abstract}
Cloud computing is a revolutionary process that has impacted the manner of using networks. It allows a high level of flexibility as Virtual Machines (VMs) run elastically workloads on physical machines in data centers. The issue of placing virtual machines (VMP) in cloud environments is an important challenge that has been thoroughly addressed, although not yet completely resolved. This article discusses the different problems that may disrupt the placement of VMs and Virtual Network Functions (VNFs), and classifies the existing solutions into five major objective functions based on multiple performance metrics such as energy consumption, Quality of Service, Service Level Agreement, and incurred cost. The existing solutions are also classified based on whether they adopt heuristic, deterministic, meta-heuristic or approximation algorithms. The VNF placement in 5G network is also discussed to highlight the convergence toward optimal usage of mobile services by including NFV/Software-Defined-Network technologies.  \\
\end{abstract}



\section{Introduction}\label{sect:introduction}

\subsection{Motivation and New Trends}\label{sect:Motivation/Introduction}

With the emergence of cloud computing, cloud providers tend to virtualize a range of telecom services by spreading the cloud computing technology toward end users and delivering mobile user's connectivity as a cloud service. In this context, authors of \cite{6616110} propose the "Follow Me Cloud" (FMC) concept, which allows services to migrate and seamlessly follow the users' mobility. 
Accordingly, services are regularly furnished from the storage computer positions that are adequate for the prevalent placement of users as well as the present state of networks. The main idea of FMC is that services pursue the users all through their mobilizations. The "follow-me cloud" approach may be achieved throughout various mechanisms. One of the key technologies is virtualization that gives the possibility to conveniently move a Virtual Machine from specific host to another without turning it off, therefore, this can offer a dynamicity on VM placement optimization with negligible impact on performance. In the Network Function Virtualization (NFV) context, we may add or remove virtual machine instances at any moment, in an unexpected manner, depending on the service chaining associated with the customer profiles. This dynamics, despite its numerous benefits, may result in sub-optimal or unstable configurations of the virtual networks. Unfortunately, the most current research work overlook this dynamicity of VM placement requests which is often managed by cloud infrastructure controllers. Moreover, VMs may experience some fluctuations within the resource utilization (e.g., a mobile application server and a web server may possess identical patterns of incoming workload while using the same CPU). \\

However, it is crucial to have the ability to precise the application requirements of virtual machine placement and already know the status of each server to properly define the constraints required for modeling a reasonable configuration. Therefore, it is highly important to identify which services are stateless, and which ones are stateful in order to ensure good performance. Hence, the critical issue lies beneath finding the optimal VM allocation and packing to reduce the number of real nodes in order that datacenter (DC) administrator could shut down the idle nodes and diminish the number of migrations while meeting the Service Level Agreement (SLA) defined with users and avoiding non-viable placement that could lead to performance degradation. This paper first focuses on presenting the major issues relevant to the placement of VMs in cloud environments. It then presents the existing optimization approaches and classifies them according to their objective functions.

\subsection{Cloud Computing: on-Demand Networking}\label{sect:Motivation-Introduction}
Cloud computing enables users to consume an on-demand computing resources, such as storage, servers, applications and networks, as instances (VMs) instead of building physical infrastructure. These resources may be swiftly provided and managed with minimal efforts 
by cloud providers. Cloud computing provides numerous exciting advantages for companies and ultimate consumers like on-demand provisioning of virtual resources, self-service ability, high elasticity, flexibility, scalability, broad network access and resource pooling. Virtualization technologies allow users to box their required computing resources into VMs. A virtual machine placement algorithm is used to define the locations of these VMs at suitable computing centers. As it will be discussed in Section II, many issues are considered for the placement of VMs: energy consumption, cost, SLAs, and load. For each issue, different solutions are proposed across several survey and research articles where each one treats the  in its unique way proposing different techniques based on various algorithms as deterministic, heuristic or approximation algorithms. Our paper aims at finding the optimal and most effective Virtual Machine Placement (VMP) approach among all existing solutions $i)$ by determining the different problems that may arise in VMP and $ii)$ by classifying the proposed solutions according to their adopted approach/objective. \\

This survey is organized as follows. Section II introduces issues relevant to VMP. The existing VMP solutions are then classified into five objective functions: 1) energy consumption minimization, 2) cost optimization, 3) network traffic maximization, 4) resource utilization, and 5) performance maximization. Each class of VMP solutions is introduced in a separate section (i.e., Sections III - VII). In Section VIII, some concluding remarks are drawn and open research problems are highlighted. 

\section{Virtual Machine Placement}\label{sect:}

\subsection{Placement Issues}\label{sect:Virtual Machine Placement Problem}

VMP is an important problem in cloud computing. It concerns the mapping between the physical and virtual machines with the objective to maximize the usage of available resources \cite{Hyser2007AutonomicVM}. Indeed, VMP is known as the procedure of choosing the most appropriate host where the virtual machine need to be deployed. The ability to migrate from a physical host to another made it possible to explore different strategies of VMP according to the constraints expressed in SLA to match different workloads \cite{Calcavecchia:2012:VPS:2353730.2353807}\cite{5990687}.\\

Large-scale cloud systems incur high cost for cloud owners, particularly in terms of energy consumption: data centers are still the major contributor of global CO$_2$ emissions of IT services \cite{6477661}. Furthermore, inefficient usage of computing resources (i.e., CPU, memory, storage and bandwidth) may result in high-energy consumption. Besides, unneeded virtual machine migrations cause additional management cost \cite{Fu2015} due to VM reconfiguration, destruction or creation of VMs and on-line VM migration, that further generate high energy consumption \cite{Hwang:2013:HVM:2514940.2515012}. Another issue pertains to performance degradation, since aggressive virtual consolidation can lead to performance degradation; therefore, One has to strike a balance of acquirable resources utilization to avoid possible degradation of performance \cite{6753800}. In this vein, a set of papers \cite{Cao2014}, \cite{Zhang2014} focuse on dealing with the compromise of minimizing the energy consumption and achieving high performance while maintaining a low-level SLA violation.\\

Along with the prominent cloud-based services, cloud data centers are affected by both spatio-temporal alteration of end users' demands and restricted available resources. Hence, the problem resides on how data centers can satisfy large number of requests for virtual resources under limitations relevant to both computation resources and link capacities \cite{6710563} \cite{6973776}.  In the same regard, the problem of resource wastage can be caused when multiple VMs are unnecessarily launched on a large number of PMs. Furthermore, the unbalanced exploitation of physical machines introduce a waste of computing resources and this may impact the placement of VMs \cite{6701467} \cite{5724828} \cite{GAO20131230}. \\

In cloud computing environment, the data centers are placed while maintaining certain geographic or rational distances among them. This may cause critical issues related to data transfer time and network traffic between data centers \cite{5662521}\cite{6168391}. The problem of virtual machine placement through a huge number of exhaustive issues is studied in \cite{ILKHECHI2015508} where A.R. Ilkechi et al. focus their research on network-aware VMP with multiple traffic-intensive components. This solution will be introduced with further details in Section IV. In the same way, numerous applications running in the cloud demand high networking resources such as intense bandwidth requirements. In this regard, R.Cohen et al. \cite{6566794} focus on VM placement problem and propose proficient manners of assigning VMs of applications, characterized by intense bandwidth, into new data centers. In \cite{6466665}, D.S.Dias et al. also reflect the traffic congestion, the connection disruption and network links in their proposed VMP algorithms. The network scalability of modern data centers is another critical issue of VMP \cite{5461930}.\\

The demand for cloud services has significantly increased along with advances in the IT industry \cite{6253575}. Data centers that house all hardware equipment (e.g., servers, network devices, power and cooling systems) and support online applications \cite{FANG2013179} require a tremendous amount of energy to process data of hosted services, resulting in huge operational costs \cite{LI20131222}\cite{5763426}\cite{5724828}. Furthermore, the electricity cost incurred by large-scale data centers is enormous since cloud service providers frequently pay per the energy consumed and the power used. \cite{6114417}. W. Shi et al. \cite{6701467} concentrate on the impact of performance, revenue and energy cost, i.e. the higher the performance levels are supplied, the higher the revenues and therefore the higher the energy costs. To cope with this issue, they develop a Multi-level Generalized Assignment algorithm for augmenting the revenue under a limited power budget and SLAs (Section V). The impact of VMP on other metrics relevant to availability, Quality of Service (QoS), and resource interference is also investigated in other research work such as \cite{6332041}\cite{6622863}\cite{6297100}.
 
\subsection{Classification of existing VMP solutions}\label{sect:Virtual Machine Placement Problem}

This section presents a global classification of various VMP optimization approaches proposed in the literature. On one hand, the determined optimization alternatives can be divided into two categories :  mono-objective or multi-objectives optimization problems (Table I). On the other hand, the selection of a possible VMP solution depends on several criteria, as discussed earlier in sub-section II-A, which vary according to several conditions and constraints. This results in a heterogeneity of possible formulations to cope with virtual machine placement problem. Taking into account the criteria diversity, we categorize the VMP solutions by purpose, i.e. those having the same goal, ranking them into five objective functions group as shown in Table II.\\

\renewcommand{\arraystretch}{2} 
{\setlength{\tabcolsep}{0.5cm} 
\begin{table*}[!h]
\begin{center}
\begin{tabular}{|p{4cm}|p{10cm}|}
  \hline
   \cellcolor{lightgray}Optimization Approach & \cellcolor{lightgray}References  \\
  \hline
 Mono-Objective  &  \cite{6503614}, \cite{Cao2014}, \cite{Fu2015}, \cite{Tang2015}, \cite{6799695}, \cite{DBLP:journals/corr/abs-1011-5064}, \cite{5488431}, \cite{6296853}, \cite{6726449},  \cite{7387769}, \cite{LI20131222}, \cite{6221099}, \cite{6477661}, \cite{6575260}, \cite{5961722}, \cite{5578331}, \cite{6332041}, \cite{6710563}, \cite{doi:10.1080/10798587.2016.1152775}, \cite{6123491}, \cite{5394134}, \cite{Hyser2007AutonomicVM}, \cite{6848063}, \cite{6566794}, \cite{6466665}, \cite{5990687}, \cite{Ortigoza2016DynamicEF}   
 \\
 \hline
 Multi-Objective &  \cite{6253575}, \cite{Hwang:2013:HVM:2514940.2515012}, \cite{5724828}, \cite{6296866}, \cite{6809418}, \cite{6679894}, \cite{GAO20131230}, \cite{5071526}, \cite{ZHENG201695}, \cite{7416960}, \cite{6701467}, \cite{Song2014}, \cite{6114417}, \cite{7179432},  \cite{ILKHECHI2015508}, \cite{5662521}, \cite{FANG2013179}, \cite{6297100}, \cite{6952725}, \cite{Pires:2013:MVM:2588611.2588692}, \cite{Anand:2013:VMP:2568486.2568500}, \cite{6753800}, \cite{6973776}, \cite{6168391}, \cite{Kakadia:2013:NVM:2534695.2534702}, \cite{DONG201462}, \cite{TORDSSON2012358}, \cite{iet:/content/books/10.1049/pbte070e_ch4}
\\
\hline
\end{tabular}
\caption{Global Classification of VMP solutions.}
\end{center}
\end{table*}

\begin{table*}[!h]
\begin{center}
\begin{tabular}{|p{4cm}|p{10cm}|}
\hline
\cellcolor{lightgray}Optimization objective & \cellcolor{lightgray}References   \\
\hline
Energy Consumption Minimization   & \cite{6503614}, \cite{6253575}, \cite{Cao2014}, \cite{Fu2015}, \cite{Tang2015}, \cite{6799695}, \cite{6726449}, \cite{LI20131222}, \cite{Anand:2013:VMP:2568486.2568500}, \cite{6296866}, \cite{DBLP:journals/corr/abs-1011-5064}, \cite{6809418}, 
 \cite{6753800}, \cite{6477661}, \cite{GAO20131230}, \cite{6973776}, \cite{6575260}, \cite{6679894},  
 \cite{6296853}, \cite{5724828}, \cite{6221099}, \cite{5071526}, \cite{GAO20131230},  \cite{ZHENG201695}, \cite{Hwang:2013:HVM:2514940.2515012}, \cite{5488431}, \cite{Pires:2013:MVM:2588611.2588692}, \cite{100000}
									\\
\hline
 Cost Optimization  & \cite{6253575}, \cite{6114417}, \cite{FANG2013179}, \cite{6701467}, \cite{6123491}, \cite{Pires:2013:MVM:2588611.2588692}, \cite{7179432}, \cite{5394134}, \cite{Hyser2007AutonomicVM}, \cite{5724828}, \cite{6952725}
               \\
 \hline
 Network Traffic Maximization &  \cite{5461930}, \cite{6753800}, \cite{6848063}, \cite{ILKHECHI2015508}, \cite{6566794}, \cite{6466665}, \cite{6296866}, \cite{6809418}, \cite{FANG2013179}, \cite{5662521}, \cite{6168391}, \cite{Kakadia:2013:NVM:2534695.2534702}
\cite{DONG201462}, \cite{6679894}, \cite{Pires:2013:MVM:2588611.2588692}, \cite{6297100}, \cite{6952725}, \cite{iet:/content/books/10.1049/pbte070e_ch4}, \cite{7461481}, 71]
           					   \\
 \hline
Resource Utilization  & \cite{Hwang:2013:HVM:2514940.2515012}, \cite{Song2014}, \cite{6973776}, \cite{6701467}, \cite{GAO20131230}, \cite{Ortigoza2016DynamicEF}
\cite{5724828}, \cite{5990687}, \cite{6710563}, \cite{7179432}, \cite{6168391}, \cite{ZHENG201695}, \cite{Ortigoza2016DynamicEF}
               \\
 \hline
 QoS Maximization  & \cite{Anand:2013:VMP:2568486.2568500}, \cite{5578331}, \cite{5961722}, \cite{6332041}, \cite{6297100}, \cite{TORDSSON2012358}, \cite{5071526}, \cite{ILKHECHI2015508}, \cite{5662521}, \cite{Kakadia:2013:NVM:2534695.2534702}
\cite{doi:10.1080/10798587.2016.1152775}, \cite{iet:/content/books/10.1049/pbte070e_ch4}, \cite{DONG201462}, \cite{7387769}, \cite{7416960}, \cite{7414158}
                \\
 \hline
\end{tabular}
\caption{Objective function-based classification of VMP solutions.}
\end{center}
\end{table*}

The technical solutions to solve VMP problems through objective functions are classified into heuristics, meta-heuristics, deterministic, and  approximation algorithms. The
mentioned solutions along with their objectives/approaches are detailed in Table III.

\section{Energy Consumption Minimization}\label{sect: III}

The majority of VMP algorithms focus on minimizing the energy consumption. Despite the fact they target the same objective, each VMP solution adopts a different approach; i.e., some minimize the DC power, others reduce the number of PMs turned on, and others focus on minimizing the network power consumption. Each approach is separately analyzed in the following subsections.

\subsection{Reducing the Energy Consumption}\label{sect:Energy consumption minimization}
Recently, minimizing the energy consumption in data centers has become the main focus of cloud providers. The existing VMP schemes aim to optimize the utilization of physical server resources or network resources. Conversely, D.Huang et al. \cite{6503614} propose an energy-aware scheme for VMP in DCs. The scheme jointly addresses the energy consumption of both network and server, taking into account the challenges encountered in terms of server capacity or multi-layered reliances of applications. The proposed approach is treated as a multi-criteria optimization problem allowing to synchronously reduce the total wastage of resources (maximizing the utilization of PMs) and the  costs of communications. A fuzzy logic algorithm is recommended to handle this issue, which provides an efficient way to balance the conflicting goals. Simulation results show the effectiveness of the suggested scheme, featured by energy saving, compared to existing methods such as bin packing algorithm and random placement. In the same way, M.Tang et al. \cite{Tang2015} propose a genetic algorithm that support the energy consumption of both the communication network and the physical servers in a DC. Then to enhance the efficiency and performance of their approach, they present a hybrid Greedy Algorithm (HGA) or memetic algorithm by incorporating : (1) an inescapable solution repair procedure which transform those infeasible in feasible ones and resolve all violation constraints (as CPU violation) ; (2) the local optimization approach that employs a heuristic mechanism to minimize the number of PMs in the VMP thereby reducing the consumed energy on the PMs.\\

In the same context of cutting down the total energy consumption, authors in \cite{6253575} suggest placing numerous VM clusters on diverse servers and thus spreading out the arriving applications on these clones of VMs which lead to increase the service reliability. Hadi et.al \cite{6253575} propose operating all these copies to serve the incoming demands. The raised problem lies on a complex resource allocation algorithm named (MERA) "Multi-dimensional Energy-efficient Resource Allocation". Hence, a heuristic algorithm is suggested to solve this issue by using Local Search (LS) and Dynamic Programming (DP) methods where LS attempts to reduce the cost of energy by shutting down the underutilized servers while DP initially identifies the number of  VM clones to be placed on servers. In other side, providing a high quality of experience also requires cloud providers to minimize the energy consumption with a low Service Level Agreement violations in DC. To accomplish this goal, Z. Cao et al. \cite{Cao2014} propose a heuristic algorithm based on tradeoff between low energy and high performance to consolidate VMs on cloud Data Centers. This framework is based on two important contributions, the first aim to classify the host overload in terms of SLA violation in two types OverS (Over SLA violation) and OverNS  through SLA violation decision algorithm (SLAVDA), whereas the second propose a "minimum power and maximum utilization" policy (MPMU) to catch a convenient placement for VM migration by increasing the minimum power policy (MP). According to simulation results with CloudSim toolkit, this framework accomplish a better energy-performance tradeoff, comparing it to older works, it performs an energy consumption minimization of 21-34\% percentage, a decrease in energy-performance of 90\% and a reduction in SLA violation of 84-92\%.\\

Dynamic consolidation is an efficient technique to minimize the energy consumption and enhance physical resource utilization. The problem resides in migrating VMs from overloaded host, which causes extra energy consumption, SLA violations and affects the migration time. To attract users, cloud Providers should achieve a high quality of service with lowest cost by simultaneously reducing the energy consumption and meeting SLAs. The main objective is to estimate energy efficiency policies based on SLA violation, energy consumption and VM migration time. The complication arises when the CPU utilization exceeds the thresholds where the necessity to invoke VM selection and VM placement procedures. In one hand, Xiong et al. \cite{Fu2015} propose a novel VM selection policy named meets performance (MP) which takes into account the resource satisfaction intensity in order to accomplish the three aspects mentioned previously. In the other hand, a VM placement technique denominated MCC (Minimum coefficient correlation) is introduced to find the suitable host featuring the minimum correlation coefficient where the migrated VM would be placed; since it helps averting performance deterioration in other VMs. simulation results prove that the suggested policies (MP, MCC) accomplish greater performance compared to existing ones regarding the three features.\\

In cloud computing environment, data centers desire enough bandwidth to achieve quality of communication between network functions and network components; hence, the occurred issue is how to save energy consumption and provide bandwidth simultaneously. To deal with this problem, S. Wang et al. \cite{6799695} propose an enhanced VMP named EQVMP (Energy efficient and Quality of Service VMP) based on computing the minimum server power and decreasing the total delay of network while preserving QoS. EQVMP is a multi-objective approach that incorporates three key features : saving energy, hop reduction and load balancing. Saving energy minimizes power on servers without SLA violation and seek the suitable placement of energy-efficient multi resources so that each VM meets its requirement by using the Best Fit Decreasing (BFD) Algorithm; hop reduction partitions VMs into segments to reduce their network traffic intensity; Load balancing aims to ensure transmission without blockage or congestion and update VM placement periodically.

\subsection{Minimizing the number of PMs}\label{sect:Energy consumption minimization}

With the advent of cloud computing, computing resources are provisioned as on-demand services over networks destined to meet user needs in a cost effective manner. 
Virtualization techniques are adopted to facilitate the usage of hardware by placing VMs on PMs to satisfy the user demands. However, virtual machine placement in large-scale environment still a very big challenge that must be tackled due to the high use of VMs which enhance the number of physical machines and then improve the power consumption. To cope with this issue, Umesh et.al \cite{DBLP:journals/corr/abs-1011-5064} present an optimal technique for VMP toward minimizing the number of required nodes. Therefore, two procedures are provided, the first is supported by linear programming (LP) while the second is based on quadratic programming (QP), these approaches contribute optimal solutions of VBP (Vector bin packing) problem by exploiting the polynomial-time solvability. According to \cite{DBLP:journals/corr/abs-1011-5064}, experiments demonstrate that "PACKINGVECTORS" algorithm provides an optimal placement of VM. \\

Moreover, Fumio et al. \cite{5488431} propose a duplicate configuration of hosting clusters for various applications employing VMs, which aim to reduce the requested number of VMs in agreement with the performance necessities of any k host failures for online applications. The problem of setting up the virtual machine is presented as an optimization problem and relies on some elements as the ability of hosting server, the requisite fault-tolerant level k, the total number of applications and their effectiveness exigencies; hence, an algorithm for VMP is defined to reach k fault-tolerance according to the specified conditions, the redundant placement allows a hosting server configuration that allows a better reliability when it comes to servers failures. Another issue for VMP is loss of performance provided by VM migration depending on type of resources and applications adopted to decrease the number of migrations and the performance failures. In this context, A. Anand et al. \cite{Anand:2013:VMP:2568486.2568500} reformulate the provisioning algorithms, i.e. ILP and FFD, by including migration overhead in placement decisions, thus aim to reduce both the number of migration and the number of hosts used and further avoid performance loss. \\

In \cite{6726449}, to bypass VM Consolidation problem (VMC), B. C. Ribas et al. propose an artificial intelligence consolidation rest on a new pseudo-Boolean Formulation of VMC problem (PBFVMC). The main objective of this issue is to employ k VMs inside N hardware with reducing the number of active physical resources simultaneously. PBFVMC focuses on removing non-linear constraints, equal constraints and reducing the number of variables, as a result, it can recognize  4 times augmentation in the size of hardware bigger than previous formulation. With this approach, the solvers spend running time on optimizing the formula whereas previous ones waste all the time to verify whether the method is suitable or not. Also \cite{LI20131222} focuses on minimizing the total energy consumption by reducing the number of running PMs. Hence, to increase the resource usage, the size of resource fragments should be minimized as well as their number.  Consequently, VMP must be achieved in a resource-balances way since the overloaded use of resources is the predominant cause of resource fragments. X. Li et al.\cite{LI20131222} define the resource utilization of each PM with space partitioning process, they further introduce a VMP algorithm named EAGLE that reduce the number of running PMs, harmonize the resource usage and minimize the energy consumption. According to expensive experiments, we can obviously show the performance of EAGLE as it can optimize 15\% of energy in comparison to the First fit algorithm. In the same regard, Ms. R. S. Moorthy \cite{100000} propose a Constraint Satisfaction based Virtual Machine Placement (CSP) to minimize both the number of running PMs and the completion duration of applications for perfectly placing VMs across various physical devices. The CSP-VMP algorithm firstly selects the optimal physical machine for placing the virtual machine then automatically migrate  the virtual machine in the overloaded Physical machine to new physical machine. Experiment results demonstrate the high performance of the proposed approach in terms of maximum user satisfaction.

\subsection{Minimizing the Network Power Consumption }\label{sect:Energy consumption minimization}

Virtualization Technology offers a powerful way to improve Data Centers by allowing high utilization of resources in a single physical server and facilitating workload movement between servers. Although, this elasticity, allowed by virtualization concept, generates management challenges of scalability problem that must be procured and handled so that the data center manager determines where VMs will be placed and how resources would be allocated to them while harmonizing thermal distribution, reducing power consumption and increasing resource usage. As in \cite{5724828}, the objective of minimizing the power consumption forms a part of a multi-attribute decision making problem that concurrently aim to reduce the energy consumption, decrease the consumption costs and the total waste of resources. However, satisfying simultaneously all this features engender conflicting objectives. J. Xu et al. \cite{5724828} propose a Modified Genetic Algorithm (MGA) to effectively search a suitable solutions for large-scale DCs, accordingly, a fuzzy-logic is applied on this algorithm to combine the conflicting goals. Whereas, \cite{6221099} focuses just on reducing the power consumption (mono-objective) while achieving performance requirements by proposing an energy aware framework for reallocating VMs in DCs. This framework computes the efficient feasible placement of VM regarding SLA constraints (hardware, QoS, Availability of services, Additional metrics) that should be decoupled to fulfill the framework's flexibility. It depends on Constraint Programming as Choco solver and Entropy open source library. The core element of this framework is the optimizer which is able to cope with the SLA requirements, minimize the energy consumption, reduce CO2 emissions and interconnect separated DCs in a federation. \\

The test results executed in a federated cloud show the benefit of this approach to decrease the energy consumption by saving 18 \%  of energy and CO2 emissions for the test case carried out. Moreover, the scalability experiments demonstrate that devising the problem in various segments is efficient to reduce the time of finding the solution. Energy consumption is an important factor for big Data centers to sustain a considerable number of leaseholders. Authors in \cite{6753800} deals with the Time-aware VMP-Routing problem (TVPR) where each occupier requires a given number of network resources and server resources for a specified time toward finding the suitable manner to map their virtual machines and route their traffic for the sake of saving energy consumption. They formulate the TVRP problem as a MILP optimization approach based on a power utilization model whose objective reside on defining the the exhaustive power consumed in DCs by using all components (servers, switches). Moreover, regarding the NP-hard complexity of TVRP problem, a heuristic algorithm is developed to fix the optimization issue. Results demonstrates the efficiency of the purposed algorithm concerning the large DC and power consumption.

\subsection{DC power Consumption/IP and WDM layer consumption minimization }\label{sect:Energy consumption minimization}

Data Centers still the major contributor of global CO2 emissions of IT services. The crucial issue lies on minimizing the power consumption and optimizing network delay in large-scale infrastructures. In this perspective, \cite{6477661} introduce an integrated algorithm for extensive Cloud Systems where the cloudified services provided by decentralized DCs are coordinated upon federated network. Firstly, Mixed Integer Programming (MILP) is introduced to compute the optimal VMP of intra/inter networking datacenters in massive Cloud Systems by considering both local physical resources and network resources through transforming network topology into virtual architecture and allocating the computing resources in DCs. Secondly, to prove the effectiveness of this solution, it has been compared with benchmark MILP model using the principle of customer loyalty (Sending the VM request to the nearest DC with enough capacity and memory in physical hosts). The proposed holistic solution allows reducing the global energy consumption that include three components: (1) Power consumption in Data Center, (2) Power dissipation in IP layer and (3) Power expenditure in Wavelength Division Multiplexing layer. Results shows the performance of holistic solution for saving energy and achieving better fairness.

\section{Cost Optimization}\label{sect: IV}

The high cost is considered as a challenge for cloud providers due to the huge requirement of cloud computing services as well as the multiple geographically distributed data centers. However, multiple elements are related to this dilemma since the cost includes the network power cost, the energy consumption cost, the electricity cost, the total infrastructure cost and the heat dissipation cost, etc. Referring to the studied articles, a group of them (\cite{6253575} \cite{FANG2013179} \cite{LI20131222} \cite{6114417} \cite{6701467} \cite{6123491}) involve in optimizing the economical costs. This section classifies these items, owing to optimize the total cost,  according to their objective functions as presented in the following subsections. 

\subsection{Reducing Electricity Costs}

One of Virtual Machine Placement Challenges is the high electricity cost in high performance computing (HPC) clouds e.g Cloud Service Providers should pay the energy consumed and the peak power used. The arisen issue is how to minimize the electricity cost for HPC providers. Studies prove the influence of load placement policies on DC temperatures and DC cooling costs, where cloud services are implemented on several geographically distributed data centers. Kien et.al \cite{6114417} propose dynamic load distribution policies for VMP and VM migration in DCs to reduce the operating cost by taking into account  the peak power prices, energy consumption and cooling energy prices. Evaluation comparison between Dynamic Cost aware policies (CA Cost Aware Distribution, CAM Cost aware distribution with migration) and Baseline Policies (RR Round Robin, WF worst Fit, SCA) demonstrates the benefits of the proposed load placement approach in large cost saving. Hence, these policies prove the impact of cooling on total cost with its large changes in data center load and the need of pre-cooling the target DC when it is necessary to prevent overheating. As seen in section II-A, \cite{6253575} aim also to decrease the energy cost by switching off the underutilized servers.

\subsection{Network Power Costs }

Statistical analyzes carried out in \cite{FANG2013179} reveal that the power of the network represents 10 to 20 \% of the total power consumption. The heavy network requirements affects negatively the overall power cost.  W. Fang et al. \cite{FANG2013179} introduce this new metric for energy saving by proposing a novel approach called "VMPlanner". This new framework aim to optimize both the traffic flow routing and Virtual machine placement  by extinguishing the unnecessary network elements to save power consumption. The "VMPlanner" exploits the elasticity of VM migration and the alignment of traffic flow routing to reduce the use of network links and therefore minimize the network power costs. Optimization Problem formulation shows its NP-hardness, then with VMPlanner, the problem can be solved by executing three approximation algorithms : (1) BMKP, Balanced Minimum K-cut Problem, for grouping virtual machines according to traffic, (2) QAP, Quadratic Assignment Problem, for grouping VMs conforming to distance, (3) MCFP, Multi-Commodity Flow Problem for routing traffic flows between VMs to save power.

\subsection{Revenue Maximization }
Data Centers often suffer from the challenge of maximizing their profit under the huge quantity of energy costs required to perform their operations. The practical method to enhance the benefit or ROI is to minimize the operational cost by supporting the PUE (Power Usage Effectiveness), adopting the virtualization technology and improving the power efficiency. However, the efficiency of these methods is limited due to the fixed conditions. In \cite{6123491}, W.Shi et al. address this problem by probing the VMP dimension in DC for increasing the revenue without invading the service level agreement. They firstly formulate the problem as a "Multi-level Generalized Assignment Problem" (MGAP) to accomplish the critical requirement of increasing the ROI under the various constraints of power budget and SLA violation, and secondly propose a first fit heuristic algorithm to resolve it. In \cite{6701467}, maximizing the revenue is a part from a multi-objective approach next to two other objectives, load balancing maximization and resource wastage minimization, Amol C. Adamuthe et al. \cite{6701467} use Genetic algorithm and "Non-dominated Sorting Genetic" Algorithms to solve the multi-objective virtual machine placement problem. In the same way, authors in \cite{Pires:2013:MVM:2588611.2588692} handle the VMP problem by incorporating three fundamental behaviors : (1) Minimizing the network traffic,  (2) Reducing the energy consumption and (3) maximizing the economical revenue. A new MeMetic Algorithm (MMA) is proposed to achieve the multi-objective optimization features.

\subsection{Resource cost minimization }
Cloud computing has become one the most requested services regarding their several benefits including high performance, elasticity, availabilty, the low cost of services and scalability. Notably, it provides and assigns applications with computing power on a shared resource pool. Indeed, Virtualized service is characterized by its flexibility on running several VMs in the same physical machine, the capacity scaling of a VM and the live migration between hosts. However, this flexibility has an imperfection on IT manager as system management becomes more complicated. Therefore, the challenge of cloud providers is to manage automatically the virtual services while ensuring high QoS of Internet-based applications and guaranteeing a cheap costs of resource management. \cite{Hyser2007AutonomicVM} presents an autonomic virtual resource manager for service hosting platforms that aim to optimize the general utility function based on the SLA fulfillment degree and the operating costs, commonly, this paradigm is able to automatize the placement of VMs and brutalize the dynamic provisioning. The core element of system architecture management is the global decision module as it deals the two main tasks : VM provisioning and VM packing,  which are expressed as two Constraint Satisfaction Problems (CSP) manipulated by CS (Constraint Solver). Results obtained through simulation tests executed by Xen hypervisor demonstrate that the constraint programming proposition is fit to settle the problem.\\

Cloud computing support a paradigm for ultimate consumers to satisfy their needs in a cost-effective manner. Resource provisioning is a crucial issue in VMP since it prescribes how resources may be allocated. Hence, to provision resources, cloud providers may furnish two payment plans to consumers e.g prepaid plan (reservation), which is cheaper but may not meet actual demands, and pay per use plan (on-demand), used for dynamically provision resources. However, while deploying a virtual machine placement, there is a big challenge of optimizing the capacity utilization due to resource allocation cost, over-provisioning and under-provisioning problems. In order to resolve these issues, S. Chaisiri et.al \cite{5394134} present an optimal VMP to implement optimized resource provisioning operations that aim to reduce their total cost. This algorithm is exploited to take a best decision making depending on the ideal key of Stochastic Integer Programming to reserve resources from VMs to any cloud service vendors. This solution is achieved through two phases : the first illustrates the number of virtual machines contributed in reservation plan whereas the second one determines the number of VMs provided in on demand plan. According to performance evaluation, in one hand, if the number of requested VMs is accurately recognized, all instances (VMs) can be contributed in reservation plan, besides a Deterministic Interger Programming is employed to reduce the overall amount of resource reservation, on the other hand if there is an uncertainty of demands and prices, a two-phase SIP algorithm is established. As a result of simulation studies OVMP algorithm based on SIP can reach the lowest total cost.\\

Along with evolving improvement in cloud computing and their advanced virtualization techniques related to network function virtualization, many papers have been focused on cost saving under better utilization of computing resources in cloud-based mobile core networks with the objective of finding the optimal placement of VNFs within the same DC. F.Z. Yousaf et al. \cite{7355586} treat this problem and examine the cost acquired through two deployment strategies based on heuristic constraints and derivations, named Vertical Serial Deployment (VSD) and Horizontal Serial Deployment (HSD) for initial implementation of VNF/VNFC.

\subsection{Connection Cost Minimization}
As a global goal, authors in \cite{7179432}  think of formulating an optimization problem for seeking the suitable VMP to reduce the cost of communication between Virtual Machines in Network Data Center. Their primary goal is to develop a beneficial algorithms to handle VMs by formulating a VM placement optimization approach. Since, Customers demand a set of requests where each one presents the desired number of VMs, hence the challenge is to determine the PMs hosting the requested virtual machines and then building a subnet connecting these PMs. In this context, there are two major aspects, resource limitation (limited performance) of PMs, formulated as a constraint, and the connection cost in each subnetwork presented as a major objective. The major intention is to reduce the total connection costs by shortening the length of networks connecting the root node with physical machines. T. Fukunaga et al. \cite{7179432} present an approximation algorithms for each case according to model's type (centralized or distributed) and requests category (uniform or non-uniform).

\section{Network Traffic Minimization }\label{sect: V }

There are several papers that study the problem of minimizing network traffic in the cloud in a manner that they try to enhance the performance of a DC through selecting the most suitable physical machines for virtual machines. In this section, we list some previous works classified conforming to their target and interest as well as their relevance. We conclude that Network Traffic Minimization can be impacted by several factors as Data Transfer Time, Average Traffic Latency, Network Traffic and Network Performance. 

\subsection{Reducing the Average Traffic Latency}\label{sect:V}

The Scalability of Data Centers can be considered as one of the most attractive prospects in cloud environment since the bandwidth usage among VMs is increasing verry quickly with the high rate required for the communication between intensive applications in Data Center. To tackle this issue, Xiaoqiao et.al \cite{5461930} propose an optimization approach of placing VMs on host machines based on TVMPP problem to embellish the network scalability. TVMPP can be optimized by reducing the traffic average latency generated by IT infrastructure. Analysis demonstrate the NP-hardness of this proposed procedure (TVMPP), however a two-tier approximation algorithm "Cluster-and-cut" is proposed to effectively solve this problem even for the enormous sizes. This heuristic algorithm firstly splits hosts and VMs into clusters separately and secondly associates them at cluster level and then in individual level. Experiments prove that the suggested algorithm can considerably reduce the accumulated traffic and decrease the computation time by comparing to other existing mechanisms that don't care about data center networking technologies and features of traffic models.\\

In \cite{6848063}, Kuo et al. presents VM placement algorithms for a particular situation which consist of allocating VMs to Data Nodes (DNs) while minimizing the highest access time between Data nodes and Virtual machines. The "virtual machine placement for data node problem" (VMPDN) is characterized by allowing each DN to only one VM. To solve this problem, they firstly introduce a 3-approximation algorithm based on threshold technique and secondly propose a 2-approximation algorithm that subdivide the global problem into small problems,and finally compute their solutions and adopt the best one. Simulation results reveal that the 2-approximation algorithm is the optimal-efficient one capable to minimize data latency in cloud systems. Moroever, in \cite{ILKHECHI2015508}, DNs can be assigned simultaneously by multiple VMs while in \cite{6848063} each VM is booked by a single DN.\\

In Mobile Cloud Computing Environments, Internet of things (IoT) based systems using a variety of mobile devices need to operate despite of connection failure or degradation. Therefore, Mobile cloud service providers can reduce network latency by moving applications close to the user. \cite{iet:/content/books/10.1049/pbte070e_ch4} suggests two cloudlet-based architectures, hierarchical and ring technologies, which aim to accomplish the mobile users requirements. The latency delay in each architecture is modeled by continuous time Markov-chain throughout different components as user nodes, cloudlets, and the main cloud. Composers use different scenarios to compare the performance of these proposed architectures. Authors in \cite{7461481} present a mechanism to minimize the one-way delay in wireless networking by introducing an easy algorithm for selecting the ever changing data plane in a mobile network. This method pick out the gateways pursuant to the designated goals of reducing the bottleneck link load, the end-to-end path latency and the network element processing load.

\subsection{Minimizing the Data Transfer Time}\label{sect:V}

Computation resources has become a critical issue in cloud computing due to the high rate of on-demand provisions. Actual VM placement procedure mostly concentrates on ameliorating the capability and proficiency of using computing resources without taking account the network performance, this can lead to place the VM faraway the data center. Consequently, the global application performance will be influenced and it is become critical to consider the network Input/Output performance because this latter affects significantly the overall application performance. To solve this issue, Jing et.al \cite{5662521} offer a network approaches for placing and migrating VMs to minimize the time of transmitting Data between VMs (the application) and related data (the data storage). The VM placement approach aim to reduce the total data access latency complying with the condition of computing the available capacity whereas the VM migration approach is happened when the operation interval surpass the SLA limen due to variable network condition that influences the users' behavior to retrieve data and damage the pertinence achievement. The simulation made by CloudSim 2.0 show that the advised technique is reliable to execution time of each task. In the same way, K. Zamanifar et al. \cite{6168391} propose a new VMP algorithm to reduce the data transfer time by optimizing simultaneously the Virtual Machine Placement and the storage allocation frequency. 

\subsection{Minimizing the network traffic}\label{sect:V}

In recent years, data centers are been progressively hired in enterprises to run a variety of applications and provide different services over a shared infrastructure. However, service misplacement can cause several problems as network link overloads, congestion and connection disruption. Indeed, network connectivity is a major influence to any data center decision that can be tackled through VM migration by alternating the traffic matrix and re-allocating services in different way. Daniel et al. \cite{6466665} present a VMP algorithm to reallocate virtual machines in DC Server contingent on memory usage, overall CPU and the traffic matrix network. The first stage of this VMP algorithm is collecting data from VMs and DC topology (Data acquisition), the second step resides on partitioning servers with a greater level of connectivity, whereas the last one consist on clustering VMs per determining the amount of traded traffic using graph theory to handle all the virtual servers. As a result, this solution enhances the quality of network traffic and the availability of bandwidth at DC, it is able to optimize 80\% of the core traffic that will be consolidated in the edge of the network. The allocation of many VMs nearby the DC with high bandwidth exigency engender congestion problem over their shared links despite its advantage to balance links utilization. \\

In \cite{6566794}, R. Cohen et al. recognize the virtual machine placement with heavy bandwidth demands and attend to maximize the revenue of the overall traffic delivered by VMs to the root node in the DC. Their scenario is similar to a storage network with intense storage needs. The mathematical formulation model of the bandwidth-constrained VM placement optimization problem demonstrates its hardness. Consequently, to overtake this complexity, they provide two approximation algorithms for delivering a capricious weight function. The first one is "the greedy algorithm"  which gives a 3-approximate solution of partitioned flow to solve the bandwidth-constrained placement problem, and can be curved to 6-approximation by using rounding procedure, whereas the second is "The Rounding Algorithm" that present a 24-approximate integral solution based on fractional Linear Programming. The convenience of this algorithm (Rounding algorithm) is its faster execution even in larger instances. They address exclusive cases of weight functions and graph topologies by using symmetric trees and considering the revenue as a simple allocated bandwidth function, the results of numerical simulations display the efficiency of the scheduled algorithms over I/O traces exported from IBM Data center. \\

Furthermore, A.R. Ilkhechi et al. \cite{ILKHECHI2015508} concentrate on the network and traffic aware VMP with multiple traffic intensive components. Their main objective is to maximize the satisfaction metric defined as the performance of placing VM on a specific PM and associated to the global overcrowded traffic in the overall network. They introduce heuristic and greedy approaches for allocating a group of Virtual Machines into a group of Physical Machines which provide excellent solutions given the sink flow demands and the communication pattern of the VMs. B. Zhang et al.\cite{6296866} develop an adept algorithm to combine VMs on PMs through ensuring a high scalability of data centers. The VMP problem is treated as a multiple criteria function owing to i) reduce the received and transmitted traffic inside a data center and ii)  save the power cost by minimizing the number of on-line PMs. In this sense, a heuristic approach is proposed to consolidate dynamic VMs and a greedy algorithm is used to manage VM requirements.\\

In the same way of \cite{ILKHECHI2015508} and \cite{6296866}, T. Yapicioglu \cite{6809418} take into account communication pattern of VMs with the major objective to reduce the traffic between frames, minimize the networking cost and decrease the average length of the traffic while minimizing the number of network components and operative servers to economize the consumed energy. They propose a clustering algorithm to bundle VMs according to their communication rates by placing virtual machines communicating frequently in the same rack. Based on simulation results, it turns out that the VMP algorithm sensitive to network traffic yields effective results regarding a greater proportion of intra rack traffic and inferior middle number of hops for each flow compared to first-fit algorithm. Therefore authors propose three heuristic algorithms to solve this issue : (1) \textbf{Greedy Algorithm} (GA), (2) \textbf{Repeated Greedy Algorithms} (RGA), and (3) \textbf{Optimal Network Function Placement for Load Balancing Traffic} (ONPL). Simulation results demonstrate the effectiveness of the proposed algorithms.

\subsection{Maximizing The Network Performance}\label{sect:V}

Dong et al. \cite{DONG201462} propose a VMP approach based on a multiple contrained resources to improve network performance in cloud. They attempt to decrease the maximum use of network links to minimize the network congestion and balance the total dispersion of data traffic. Reducing the entire communication traffic is considered as a Quadratic Assignment Problem (QAP) that is NP-hard. To solve this combinatorial problem, Dong et al. \cite{DONG201462} propose Ant Colony Optimization Algorithm (ACO) attached with 2-opt Local Search (LS). Simulation analysis for the proposed algorithms, with different technologies as Tree, VL2 and Fat-Tree yield better optimization results compared to simulated annealing (SA), local search (LS) and clustering algorithm \cite{6679894}. As a result, Maximum Link Utilization (MLU) is reduced by 20\% and the number of links between applications is saved by 37\%. For the same purpose, Authors of \cite{6679894} suggest a VMP strategy founded by a novel two phase heuristic algorithms to reduce the network congestion through minimizing the power consumption of physical resources and network components to guarantee a satisfactory network performance. \cite{Kakadia:2013:NVM:2534695.2534702} aim also to ensure a high performance and  promise an optimal network utilization in big data centers by consolidating VMs using network awareness. D. Kakadia et al. \cite{Kakadia:2013:NVM:2534695.2534702} firstly propose "VM-Cluster formation algorithm" to group the VMs in reference to their traffic swap models and then suggest a greedy consolidation algorithm "VM-Cluster placement algorithm" for placing VM-Clusters toward centralizing traffic in the same group. This solution can save 70\% of the internal bandwidth and improve 60\% of the application performances.\\

In telecommunication domain, mobile operators need an efficient mobile cloud to meet their customer requirements thereby dealing with the intense mobile data traffic and the humble average revenue (ARPU). In this context, authors in \cite{Taleb:2013:GRA:2507924.2508000} propose a heuristic algorithm to diminish and avert the density of gateway relocations in carrier cloud thus reducing the gateway relocation cost while guaranteeing efficient placement of virtual network functions. Similarly, \cite{7248929} also lies on avoiding the relocation of mobility anchor gateways (S-GW) by placing their VNFs away from the UEs and ensuring QoE by placing "data anchor gateways" (PDN-GW) VNFs nearer to mobile devices. To achieve this two goals, the authors in \cite{Taleb:2013:GRA:2507924.2508000} propose a VNF placement algorithms based on linear programming to find an efficient placement of VNFs in both PDN-GW and S-GW for cloud carrier and that is to create elastic mobile core network conform with Virtual 5G Network Infrastructure. In the same way, \cite{7355586} propose a Fine-grained scheme based on computing reference resource affinity score RRAS values of each hosted VM for optimal management and decision of VFNs. This approach can optimize the Life Cycle Management (LCM) operations on the VNF instances and reduce the occurrences of the expensive VM management operations.

\section{Resource Utilization}\label{sect: VI}
The utilization of the lowest number of servers is an important factor to consider when searching the power efficiency in Data Centers. Consolidation of Virtual Machines (VMs) on servers implies collecting numerous virtual machines in a unique physical server by growing the resource utilization. It allows to shut off idle servers which yields a great deal of energy saving. The VM condensation is realized either in a static manner by allocating physical resources to VMs depending on the strong consumer demand (over-provision), and this produce a waste of resources since workloads are not often at peak point. Or in a dynamic manner as VM capacities change in accordance with the current workload requirements and this helps to effectively use data center resources. In this context, this section focuses on resource utilization maximization through minimizing the resource wastage, maximizing the resource usage and increasing elasticity.

\subsection{Maximizing resource usage}

To deal with scalability and energy consumption problems, F. Song et al. \cite{Song2014} propose an optimization-based algorithm for virtual machine placement approach that takes into account both server constraints and dependencies between VMs and applications levels so as to maximize the resources allocation and minimize the Data Center transmission traffic in a short duration. Their principal objective is to minimize the number of physical hosts, improve the scalability and decrease the energy consumption. Targeting the same goal, N. Trung \cite{6973776} not only pretend to increase the resource utilization but also to equilibrate the resources utilization beyond several measures beneficial to minimize the number of running servers. They propose a complex Balanced Resource Utilization (Max-BRU) algorithm characterized by multiple resource-constraint measures such as the d-th dimensional resource utilization ratio (RUd) and the resource balance ratio that allow to find the most appropriate server to deploy virtual machines in large DCs. Through in-depth simulations, Max-BRU algorithm can balance the use of resources and makes it more efficient by reducing the number of active physical servers.\\

With the emergence of energy consumption in cloud computing architectures, there is an increasing need of energy aware resource management in data centers \cite{Hwang:2013:HVM:2514940.2515012}, taking into consideration the relation between multiple kind of resources ( e.g. Bandwidth, CPU, Network usage, Disk Space, Memory size) and the VMs resource requests. I. Hwang \cite{Hwang:2013:HVM:2514940.2515012} characterize the resource demands as random variables RVs relative to two metrics : expected mean and standard deviations. Then, they formulate the problem of VM consolidation as a Multi-Capacity Stochastic Bin Packing (MCSBP) issue and suggest a heuristic algorithm (First Fit Algorithm) to solve it. This approach provides efficient results with rational resources management.

\subsection{Minimizing the resource wastage}
The unbalanced use of residual resources can impact the placement of virtual machines and cause a waste of resources \cite{5724828}. Fig. 1 depict an example of resource wastage where host has a little available memory but a lot of unused CPU capacity which prevents the host from accepting a new virtual machine due to lack of memory. Whence the need to balance the use of resources in diverse dimensions by reducing the remaining resources wasted on a server "$W$", which is defined as the summation of subtractions between the minimum controlled residual resource "$R_i$" and the other k different resources "$R_k$" (1). As seen in section (II), J. Xu et al. \cite{5724828} consider the VMPP as a multi-attribute optimization problem of minimizing simultaneously the thermal dissipation costs, the energy consumption and the total resource wastage. They propose a "modified genetic algorithm" (GA) to effectively select the suitable solutions and introduce a fuzzy-logic to incorporate the various objectives.

\begin{equation} 
	W =\sum_{{\i\neq k}}^{n}{R_i-R_k}  \quad  (1).
\end{equation}

\begin{figure}[!h]
\centering
\includegraphics[scale=0.9]{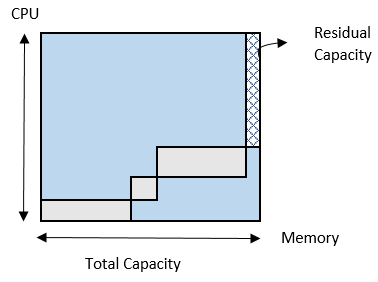} 	
\caption{Resource wastage}
\label{Resource wastage}
\end{figure}

While in \cite{6701467}, minimizing resource wastage is joint with maximizing profit and maximizing load balancing, A. C. Adamuthe et al. \cite{6701467} propose "Genetic algorithms" to solve the problem of VMP in data centers, but simulations prove that the NGSA-II give good and diversified solutions compared to simple NGSA and GA algorithms. In like manner, Y. Gao et al. \cite{GAO20131230} propose a multi-objective ACS algorithm to reduce both the aggregate energy consumption and the resource wastage based on permutation VM assignment. Analysis prove its high performance compared to multi-objective genetic algorithms as Max-Min Ant System (MMAS) algorithm and bin packing algorithm. Whereas, \cite{ZHENG201695} propose a novel solution called VMPMBBO to tackle the virtual machine consolidated placement VMcP problem. This novel approach aim to decrease simultaneously the energy consumption and the resource wastage based on the bio-geography-based optimization strategy.

\subsection{Elasticity maximization}

Taking into account the workload dynamics of modern cloud applications, elasticity is a very important issue to address for Cloud Service Providers in order to deal with under-provisioning (saturation) and over-provisioning (under-utilization) problems of cloud resources. To tackle the problem of placing VMs in cloud infrastructure management. The VMP problem should be solved dynamically to provide a common workload of modern applications. In this context, authors in \cite{Ortigoza2016DynamicEF} propose a taxonomy in order to discern all possible defiances for Cloud Service Providers (CSPs) in dynamic environments, classified with elasticity and overbooking through implementing: (1) Vertical elasticity to dynamically adjust the capacities of virtual resources inside a VM, and (2) horizontal elasticity to adjust the number of VMs.\\

Elasticity in term of resource utilization is referred to how the datacenter can appease the high requirements of VMs resources within the capacities limitations of links and physical machines \cite{6710563}. K. Li et al. \cite{6710563} focus on maximizing elasticity to deal with the restricted resources and the different users demands, they propose a ranked algorithm of virtual machine placement for a multiple layer bunch which heuristically place VMs in succession from the upper layer to the lower layer. As a result, this scheme furnish a approximatively optimal results. On the other hand \cite{5990687}, the service variability in workload may unbalance resource utilization, therefore using elastic services lead to match these workload fluctuations based on on-demand capacity allocation. This variability depends on Service Level Agreement thereby the problem lies on increasing the profit cloud provider from flexible placement to SLA in virtualized Data centers. In the same vein, \cite{5990687} cope with the variety of placement restrictions by proposing a new combinatorial optimization algorithm capable of finding the optimal solution, this problem is named Elastic Services Placement Problem (ESPP) and allows service suppliers to augment their profit through SLA conformity placement. Since ESPP is NP hard to be solved, a simple transformation is used by presenting ESPP as a multi-unit combinatorial auction, and then a delayed column generation method is implemented to acquire an efficient solution in a reasonable amount of time. This procedure submits suitable results for loading large resource pools quickly and effectively.\\

According to \cite{Ortigoza2016DynamicEF}, elasticity is classified into two types : vertical and horizontal, vertical elasticity is referred to the capability of cloud services to dynamically change resource capacities of VMs while horizontal express the ability to adjust the number of virtual machines; elasticity is an important matter to cope with under-provisioning and over-provisioning problems. J. Ortigoza \cite{Ortigoza2016DynamicEF} propose a dynamic approach based on three pertinent parameters: the VMs resources competences, the VMs number and the VMs resources usage (referred to overbooking) in order to efficiently attend applications workload and customers' requests for virtual resources.

\section{Quality of Service Maximization}\label{sect: III}

The complexity of virtual machine placement in Software Define Networking (SDN) is reflected on controller Placement Problems\cite{7511136}, given a network topology and a response time bound, how many controllers are needed, where to place them and which switches are assigned to each of the controller. Since the QoS is the initial concern of the network operators in the placement of SDN controllers, authors in \cite{7416960} propose three heuristic algorithms (incremental greedy algorithm,primal-dual-based algorithm and network-partition-based algorithm) to solve this controller placement problem with the QoS requirement. To insure a good QoS, there are many important metrics to take into consideration. These constraint are expressed in the SLA between the customer and the cloud provider. Therefore to achieve a high quality of service, performance maximization and high availability must be taken into account, this is the subject of this section.

\subsection{High availability}

Virtual Machine Placement is a critical problem that must be tackled by satisfying multiple constraints as QoS, energy conservation, performance, security, etc. The high availability of all applications running in VMs is considered as one of the most critical management domain for VM placement. Thus, if a virtual machine is noticeable as k-resilient, it will be moved to a non-failing host without moving the other virtual machines. This technique is based on a combination of live migration and Hardware Predicted Failure Analysis (HwPFA) alarms with the purpose to allow the continuous running of VMs despite host failures. The non-resilient VMP is defined to reach the resource feasibility constraint but it does not provide High availability properties where the necessity to adjoin flexibility limitations thereby derives the elastic virtual machine placement algorithm denominated RVP. The challenge in solving the NP-hard RVP problem is the second order logic constraints; consequently, to address this frustration, E. Bin et.al \cite{5961722} propose an efficient solution that aim to convert constraints of RVP probem to simpler constraints regardless the failure sequence by adding shadow VMs to the placement computation. Hence, this transformed algorithm enables a Constraint Programming (CP) scheme to provide an optimal solution in a short time. Results exhibit the convenient benefit of the proposed approach in optimizing load balancing, reusing backup in production host and ensuring high availability of evacuating VMs from failed hosts without reshuffling other VMs.\\

As cloud computing is an on-demand service, customers must be able to have constant availability of their VMs even with the variability of workloads. \cite{6332041} highlights the difference between vertical and horizontal resizing aiming to provide a cloud model with the best availability by developing Availability-Aware Placement algorithm. Vertical policy scales up the resource of existing VMs, while horizontal policy adds resource by creating new VMs. However, in this paper, the authors assumed that all the physical hosts are identical which is not true.\\

Changing trends in the telecommunications industry will reveal whether the cloud, NFV and SDN can be an effective paradigms for operators to manage automatically their operations. However, the availability gets an important issue while using the ready made data centers easily built (COTS) since it engenders high failure probabilities. Consequently in Network Function Virtualisation technology, the convenience of cloud infrastructure must be treated from the physical layer till the hypervisor layer and the flexibility technique need to be integrated into software design and service delivery \cite{7414158}. 

\subsection{Performance maximization}

Provisioning resources in cloud computing enable to improve the use of resources through placing virtual machines on a limited number of physical machines while meeting the required capacities. The capacity demands used by most applications are itemized on memory usage, storage and network bandwidth. However, the extra CPU requirement makes the capacity requirement incomplete and this may be solved through migrating VMs, which itself can cause a loss of application performance. In this vein, A. Anand \cite{Anand:2013:VMP:2568486.2568500} aim to fulfill SLAs performance by including VMM (Virtual Machine Migration) and CPU overhead constraints in VMPP to minimize the effect of application performance along with achieving the crucial goal of reducing the number of hosts. They make some changes in traditional ILP and First Fit decreasing (FDD) algorithms to evaluate the proposed constraints of CPU overhead and VMM. The ILP algorithms give optimal results but in a long duration, whereas FDD heuristics are scalable and much faster than ILP as it yields suboptimal solutions used for real time decision making. As a result, involving VM migrations and CPU overheads in the placement algorithm lead to reduce the number of migrations by more than 84\%. 
\\

Virtualization Technology has been an integral aspect of cloud computing environment. Despite its manageability and utilization advantages, the virtual machines affect negatively the utilization performance in comparison to original behavior even if those instances are executed on a particular server. \cite{5578331} focuses on the main benefits and effects of multi-core architectures that may restrict the efficiency influence of virtualization and this by examining the performance results of multicore cache system on application operating within Xen Hypervisor. Oprofile and Xenoprofile make use of the hardware performance monitors to provide information about resources consumption, current status of operating systems and applications. Hence, various strategies are used for placing virtual machines to Physical CPU (Harpertown Processor) thus are summarized in three cases according to the number of virtual machines residing on the single node ( single, two and four VM placement), all this placement approaches centralize only on second layer of cache sharing (the last level cache in Harpertown processor). The simulation results prove the high performance of placing VMs in caches in contrast to Xen default VCPU scheduler.\\

Moroever, \cite{TORDSSON2012358} proposes a cloud brokering mechanism where the virtual machines of a service are deployed across multiple clouds to maximize performance, while considering various limitations in terms of budget, load balance, service configuration, etc, thereby using scheduling algorithms applications considering the integer programming formulations and the price-performance placement tradeoffs. Cloud computing has become a very requested service due to their multiple benefits including low cost of services, scalability, elasticity, high performance and availability. Notably, it provides and assigns applications with computing power on a shared resource pool. Indeed, Virtualized service is characterized by its flexibility on running several VMs in the same physical server, scaling a virtual machine capacity and live migration between hosts. However, this flexibility has an imperfection on IT manager as system management becomes more complicated. The key challenge of cloud service providers is how to manage virtual machines automatically while ensuring high QoS of hosted applications and low resource management costs. \cite{5071526} details a dynamic placement of VMs on a set of PMs to minimize the number of migrations and the number of active physical servers toward providing an optimal configuration that takes into account SLA constraints. H. Nguyen et al. \cite{5071526} presents an involuntary virtual resource manager for service hosting platforms that aim to optimize the global utility function based on the SLA fulfillment degree and the operating expense, generally this paradigm is able to automate the dynamic placement of VMs. The core element of managing system architecture is the global decision module as it deals with two main features : VM provisioning and VM packing, which are anticipated as two Constraint Satisfaction Problems (CSP) managed by CS (Constraint Solver). Simulation results done by Xen hypervisor demonstrate that the constraint programming approach solve the optimization problem.

\subsection{Minimizing the resource interference}

Virtualization is basically a number of VMs sharing multiple resources includes memory, network, bandwidth, computing power, CPU,etc. Thus, the issue of isolation need to be dealt with \cite{Kim:2012:VMP:2401603.2401656}. In a competitive IT's market Cloud providers should minimize the SLA violations \cite{6297100} to meet the consumer's expectations. Incubator \cite{Kim:2012:VMP:2401603.2401656} is a server bundle that measures traffic dispersion and requirements of instances. The concept of placing VMs with contrasting traffic diffusion within the uniform server minimizes the over-subscription of resources and grows the use of network links. Authors in \cite{Kim:2012:VMP:2401603.2401656} propose a system that measures the time-varying traffic for tenants VMs for a month, and then use a stable marriage algorithm to place these VMs to reduce network over-subscription. \cite{6297100} investigates interference-aware VMP (IAVMP) problem to ensure the quality of service exigencies of user subsciptions and efficiently maximize the network I/O performance. Authors formulate IAVMP by Integer linear programing but due to its complexity, a polynomial heuristic algorithm is proposed to tackle this problem which guarantee a high performance compared to other VMP algorithms.

\subsection{Reliability}

Reliability is an important aspect to guarantee the high Quality of Service, regarding the number of running VMs in cloud data centers; it is arduous for cloud services to satisfy VMs performance due to software and hardware challenges that cause VM failures. Therefore, enhancing the reliability is a challenging issue to address Virtual Machine Placement problems. A. Zhou et al. \cite{7387769} propose a new recurrent VMP algorithm to improve the accuracy of cloud services. This approach called OPVMP (\textit{optimal redundant virtual machine placement}) aim to reduce the lost time and the network resource consumption under fault-tolerant requirements. It is based on tree-step process where each one is defined with an algorithm, (1) efficient VMP, (2) host server election and (3) recovery strategy agreement, using a heuristic approach is able to select the suitable host servers and to determine the optimal VM placement.\\

In the same way of ensuring a high reliability of cloud applications, authors in \cite{doi:10.1080/10798587.2016.1152775} propose a VM placement scheme subject to flexibly choose the fault-tolerant methodology (SelfAdaptionFTPlace) of cloud applications. The system architecture of SelfAdaptionFTPlace lies on three stages : (1) convert the application demands to constraint patterns, (2) select the flexible fault-tolerant procedures, and (3) solve VMP problem. The constraint model consider three factors : resource consumption, failure rate and response time; thus a two-phase approach is adopted to settle this problem, where the first phase consist of solving the fault-tolerant strategy of VMP considering the uniformly changing of cloud applications hindrance parameters. While the second one allows you to solve the VMP paradigm regarding the solution of the first stage. Results demonstrate that SelfAdaptionFTPlace obtains better performance compared to existing methods (RandomFTPlace, NOFTPlace and ResourceFTPlace).

\section{Optimal VNF placement in 5G Network}
Telecom world is changing exponentially to become virtualized and cloudified. "Software-Defined Networks" (SDN), "Network Function Virtualization" (NFV) and Mobile Cloud Edge Computing are the key elements of a global economic trend toward 5G. The technology of the fifth generation network is divided into four areas: 
\\

\noindent \textbf{Network softwarization}: Network Softwarization is a general revolution tendency for deploying, configuring and updating network functions and/or equipment beyond software programming. This virtualized network functions are dynamically allocated in cloud computing considering the software reliability and the storage capabilities.
\\

\noindent \textbf{Network management/Orchestration}: NFV and SDN technologies constitute the foundation for managing the life cycle of logically isolated network partitions, called "slices". When creating a slice, the management and orchestration functions, NFV-MANO, will provide primary capabilities: select functions requested, launch them on a virtualization platform, and connect them via virtual networks created on physical infrastructure. 
\\

\noindent \textbf{Fronthaul/backhaul} : while requirements of 5G will be set until 2020, capacity and coverage are the most important that must be emphasized in every evolution in cellular network. To achieve this requirements, mobile operators will turn to small cells through the addition of the Remote radio heads (RRHs) operated with baseband units (BBUs). Mobile fronthaul (MFH) is a transport network connecting RRHs to BBUs and mobile backhaul (MBH) is a transport network connecting BBUs with core network functions, such as MME, S-GW/P-GW, .etc. 
\\

\noindent \textbf{Multi-access edge computing} : provides service capacities, application developers and networking infrastructure based on virtualization technologies. The goal is to define a set of APIs that enable the creation of virtual network functions (VNFs) that respond to all the needs of a mobile communications network, including security, orchestration and portability to deliver ultra-low latency and high bandwidth of applications.

\subsection{VNF placement Issues and Challenges} 

The compelling deployment of network function virtualization requires multi objective functions, such as minimizing CAPEX and OPEX, reducing the network latency and decreasing the number of active nodes in the network. However NFV placement with considering the tradeoffs between this objectives may cause many conflicts since hosting multiple VNFs in the same hardware can lead to scalability issues \cite{7608294} for example minimizing the number of active node may growth the aggregation traffic in physical links which affect the network latency, while reducing network latency may be cost-effective due to the redundancy of resources to deploy VNFs. Three Integer Linear Programming (ILP) models are proposed in \cite{7608294} to solve VNF placement problem with VNF service chaining while ensuring resiliency against single link, single node and single node/link failures.
The diverse network functions forming 5G infrastructure are placed on VMs that can be switched among various PM (Physical machines or servers).Then the power consumption can be minimized by shutting down unused resources, however, we don't know exactly the resources that network function requires and if  placing more virtual network functions in fewer physical resources may deteriorate user's experience of the service and violate SLAs.  \\

The challenge of placing virtual network functions (VNF) over service chains is studied in \cite{7859379} to guarantee a traffic and energy-aware cost reduction thereby proposing an algorithm that combines a "sampling-based Markov approximation" (MA) with matching theory to find an efficient solution of saving functional costs and network traffic costs. This algorithm is named SAMA , it's based on two phases : i) identify the nodes where vNFs may be implemented and ii) place vNFs in a manner to reduce the overall cost provoked in the structure. This program minimizes the state space of possible alternatives that immediately influences the convergence duration. 
In order to support increasing traffic, mobile operators will need to introduce a number of small cells through the addition of base stations or remote radio heads (RRHs) operated with baseband units (BBUs) and connected by Mobile Fronthaul/Backhaul. The consolidation of mobile backhaul and fronthaul inside a single carrier network builds a new architecture named 5G crosshaul. This anticipated structure for the 5G networks \cite{pub2665423} requests a completely incorporated management of network resources in an elastic, scalable and cost-effective manner by integrating an SDN/NFV control framework. \\

The VNF placement objective is to reduce the resource expenditure and improve the trustworthiness. Authors in \cite{pub2665423} propose two key SDN/NFV technologies addressing energy utilization and cost-effective resources : "the Energy Management and Monitoring Application" (EMMA) and "the Resource Management Application" (RMA) with the intention to enhance performance and decrease costs. The conspicuous goal of EMMA is to minimize the power consumption of millimeter wave mesh 5G network by turning off the mmWave components characterized by limited traffic requirement. Meanwhile the goal of RMA is to manage Crosshaul resources and maximize the resource utilization in cost-efficient way and also endure the variable demand of 5G Points of Attachment (5G PoA). Fig. 2 taken from \cite{pub2665423} shows the physical infrastructure of which Xhaul is split into three differentiated layers.\\

The modeling work of \cite{Cao2017} for VNF placement optimization problem involve multiple optimization objectives to achieve minimal bandwidth dissipation and the smallest maximum link application simultaneously, four genetic algorithms have been proposed by using the frameworks of two existing algorithms MOGA (multiple objective genetic algorithm) and NSGA-II (non-dominated sorting genetic algorithm) : Greedy MOGA, Greedy NSGA-II, Random MOGA and Random NSGA-II ) . Numerical Analysis show that Greedy-NSGA-II accomplishes high performance compared to the four proposed algorithms. The critical issue of the mobile operators sustainability in terms of 5G relies on providing high service performance and immense data rate connectivity. The system scalability should be flexible to admit large amount of mobile applications.\\

Taking into account this requirements,as example for video streaming service where customers are attached to the network to hear video streams, they usually stop depleting resources. The service will be impacted in terms of play-out starvations, packet loss and startup delay due to the lack of static allocated resources (i.e, physical memory, CPU, cache, buffer, swap) that are insufficient to manipulate the number of connections wish degrade user's QoE. However if the resources are over-provisioned, this will increase energy consumption and diminish revenue. To overcome this limitations, \cite{7511377} proposes a virtual infrastructure that dynamically expands or reduce the resources to provide an optimal usage of resource and ensure a good QoE of the provided services. 

\subsection{Toward 5G Slicing}
Regarding the virtual machine placement problems already discussed, one need to understand how Telecom Providers will bypass the issues related to virtual network function placement in 5G network. In this context, network slicing offers a number of significant advantages that are particularly useful for the conception of NGN (next generation networks) \cite{7926923}. Slicing provide flexible VNF placement that may enhance network performance and reduce operating costs. It addresses the deployment of several logical networks as separated business operations on a common physical infrastructure \cite{9d560172e1ba408fb355946fb8627734}. With the tremendous growth of cloud-based technologies towards integrated 5G infrastructures, diverse architectures have been already proposed. Such propositions are presented for example in \cite{Nikaein:2015:NSE:2795381.2795390},  \cite{9d560172e1ba408fb355946fb8627734}, \cite{7503760} which provides the means to support the expected service diversity, flexible deployments, and network slicing. Authors in \cite{Nikaein:2015:NSE:2795381.2795390} display the pattern of 5G-ready architecture and a NFV-based Network Store along with a network slicing for 5G applications. The objective of the proposed network store is to produce programmable VNFs and furnish 5G slice that perfectly matches end user's demands. Meanwhile network splitting aims to exploit virtual networks in physical infrastructure by isolating the virtual resources and ensuring high performance of virtual networks.\\

In the same way, authors in \cite{7503760} deliver New Architectural Design to Open Cloud-Based 5G Communications. The proposed architectural pattern in \cite{7503760} depends on considering network slicing as a brain wave in cloud-based RANs owing to increase the scalability of current RAN systems. Based on recent works \cite{Nikaein:2015:NSE:2795381.2795390}, \cite{7503760}, the Network Slices Architecture illustrated in fig. 3 is splitted in 3 layers:\\
\begin{figure*}
\begin{center}
\includegraphics[width=\textwidth]{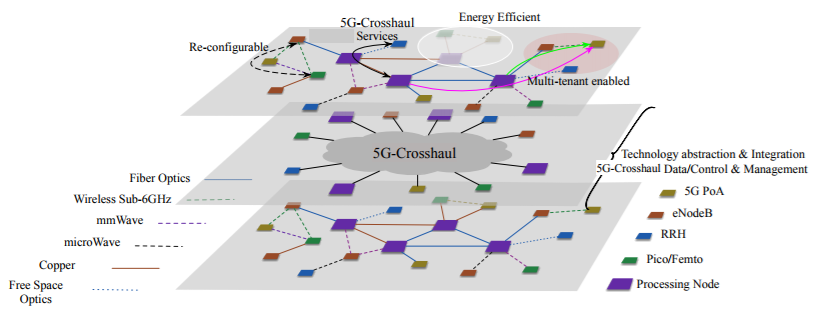}
\end{center}
\caption{5G crosshaul physical architecture}
\label{5G crosshaul physical architecture}
\end{figure*}

\noindent\textbf{The business layer} is an Application and Network Functions Market place used to provision diverse use-cases(e.g, high mobility, speed, IOT). It establishes a slice that encrypt all the requested informations from 
the service layer to create the desired service.
\\
\noindent\textbf{The service layer} sustain the configuration, management and scaling of the services operational bundle regarding their specific use-case requirements defined in the "slice manifest". Based on decision making, it accomplishes network life-cycle service management and has a direct access to network informations requested by VNA.
\\
\noindent\textbf{The infrastructure layer} maintains the reconfigurable cloud ecological system in a real time and uses virtualization for high speed services.
\begin{figure*}
\centering
\includegraphics[width=\textwidth]{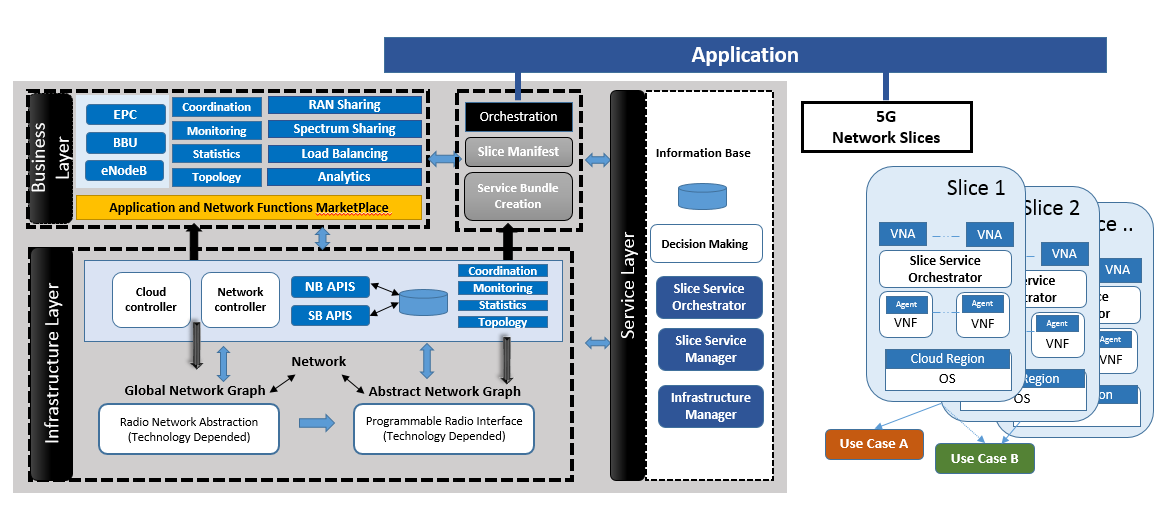}
\caption{5G Network Slicing architectures (\cite{Nikaein:2015:NSE:2795381.2795390}, \cite{7503760}) }
\label{5G Network Slicing architectures }
\end{figure*}

Network Slicing will become more relevant and more important in the context of 5G Networks. It is created to deliver services to many vertical industries, for example both SDN and virtualization NFV are actually creating programmable network as a service that will expose certain network capabilities to the business layer, then those services can be defined to different vertical industries. Particularly, some industries will need very low latency and high security requirements while others will require high bandwidth, which means each slice will have different characteristics. However, the user experience will be the same, since network slices are isolated from each other in both user plane and control plane. \\

According to a white paper published by Ericsson, future 5G networks will have a pliable structure where network slices allocate separately capacity, speed and coverage requirements. Hence, with network slicing technology, a single physical network can be splitted into multiple virtual networks allowing Operators to provide various services and different customer slices.

\section{Conclusion}\label{sect:}

Virtualization technologies heavily used by cloud computing environments enable virtual machines (VMs) to be transferred between physical systems. In such a competitive field, cloud providers seek to figure out new ways to  acquire optimal virtual machine placement by addressing various problems as energy consumption, high cost, performance degradation, SLA violation,etc. According to previous researches, overall there are two types of solutions : multi-objective functions and mono-objective functions, particularly we classified those solutions into five objective functions as follows : (1) Minimizing the  Energy consumption, (2) cost optimization, (3) Network traffic minimization, (4) balancing Resource utilization and (5) ensuring high quality of service.  Various protocols, heuristics, algorithms and architectures were surveyed. Virtual Machine Placement algorithms has been solved as heuristic, meta-heuristic, deterministic and approximation algorithm depending on the angle they treated the VMP issue with. The majority of works use heuristic and/or meta-heuristic algorithms since it provides good quality solutions. Finding an optimal solution with all the constraints facing VM and VNF Placement in 5G networks would be the subject of our future work.

\newpage
\begin{*}
\centering


\section*{Supplementary material}

| Technical Solutions | Algorithms | | | References | Objectives |
|---|---|---|---|---|---|
| Heuristic Algorithms | Best Fit Decreasing (BFD) | | | [70] | -Hop reduction, saving energy and load balancing |
| | First Fit Decreasing (FFD) | | | [3] | -Reduce both the number of migration and the number of hosts used<br>-Avoid performance loss |
| | First Fit (FF) | | | [60] | -Reduce the operational cost<br>-Maximize ROI |
| | | | | [32] | -Reduce cost and maximize resource utilization – scalable for a number of resource types |
| | Greedy Algorithm (GA) | | | [43] | -To minimize the operating cost |
| | | | | [22] | -Reduce the network power cost |
| | | | | [34] | -Maximize satisfaction (VM performance) |
| | | | | [76] | -Improve the scalability of the DC<br>-Minimize the communication traffic within DC |
| | | | | [35] | -To improve the application performance and optimize the network usage in large data centers by consolidating VMs using network awareness |
| | | | | [4] | -Minimizing the traffic (Load balancing) and reducing operator cost |
| | | | | [65] | -Avoiding gateway relocations under efficient network function placement in carrier cloud |
| | Polynomial heuristic algorithm | | | [47] | -Reduce interference resources, high QoS |
| | Others | Energy-aware VMP | | [31] | -Minimize network data transmission<br>-Reduce Energy Consumption<br>-Minimize total resource wastage and communication costs |
| | | SLA violation decision algorithm (SLAVDA) | | [10] | -Reduce energy consumption with low SLA violations in DC |
| | | MP meet performance and MCC minimum correlation algorithm | | [24] | -Avoid SLA Violations<br>-Reduce energy consumption and VM migration times |
| | | MKR Multiple K redundant method | | [48] | -Reduce the number of PMs and performance failures |
| | | EAGLE energy-efficient online VM placement algorithm with balanced resource utilization | | [46] | -Balance the utilization of multi-dimensional resources<br>-Reduce the number of running PMs<br>-Minimize the total energy consumption |
| | | TVPR heuristic Algorithm | | [16] | -Reduce the total power consumption |
| | | TVMPP two-tier approach | | [50] | -To improve the network scalability<br>-To minimize the average traffic latency caused by network infrastructure |
| | | ROP rate optimal placement algorithm | | [75] | -Reduce the data transfer delay |
| | | VMP algorithm based on traffic matrix model | | [17] | -Improve the quality of the network traffic and the availability of bandwidth at the DC<br>-Reduce the core traffic |
| | | Hierarchical VM placement algorithm | | [44] | -Maximize elasticity to deal with the limited resources and the variation of users requests |
| | | Availability Aware Placement | | [71] | -To provide a cloud model with the best availability. |
| | | OPVMP | | [79] | -To reduce the lost time and the network resource consumption under fault-tolerant requirements -Reliability |
| Meta-heuristic Algorithms | Memetic Algorithm (Hybrid Genetic Algorithm) | | | [66] | -Resolve all constraint violations<br>-Reduce the total energy consumption (physical server+ communication network) |
| | Modified Genetic Algorithm (MGA) | | | [72] | -To minimize the total resource wastage, the power consumption and thermal dissipation costs |
| | Genetic Algorithm (GA) | | | [2] | -Maximize revenue, load balancing, minimize resource |
| | Multi-objective Memetic Algorithms (MMA) | | | [55] | - Energy consumption minimization, network traffic minimization and economical revenue maximization. |
| | Ant Colony Optimization (ACO) | | | [36] | -To improve network performance in cloud |
| | | | | [26] | -Minimizing total resource wastage and power consumption based on permutation VM assignment |
| | Biogeography-based optimization (BBO) | | | [78] | -To minimize both the resource wastage and the power consumption |
| | Column generation algorithm | | | [7] | -To maximize the providers benefit from SLA compliant |
| Deterministic Algorithms | Integer Linear Programming (ILP) | | | [5] | -Reduce the number of required nodes |
| | | | | [3] | |
| | | | | [67] | -To maximize performance |
| | | | | [30] | |

|  | Quadratic programming | [5] |  |
|---|---|---|---|
|  | Dynamic Programming | [27] | -Minimize the total energy cost in the cloud system<br>-Increase Energy efficiency |
|  | Pseudo-Boolean Optimization (PBO) | [57] | -Avoid Virtual Machine Consolidation Problem |
|  | **Constraint Programming (CP)** | [1] | -Minimize both the number of running PMs and the completion time of the applications |
|  |  | [19] | -To decrease the energy consumption by saving the amount of energy and $CO_2$ emissions |
|  |  | [33] | -High QoS of hosted applications and low resource management costs |
|  |  | [29] | -Maximize and balance the resource utilization |
|  |  | [6] | -High availability of evacuating VMs from failing hosts<br>-Achieving load balancing optimization |
|  |  | [68] | -Minimizing the number of active physical servers and the number of migrations to |
|  |  | [12] | -To guarantee the reliability of cloud applications |
|  | **Mixed Integer Linear Programming (MILP)** | [37] | -Minimize the global energy consumption that include power consumption in DC, IP layer and WDM layer |
|  | **Stochastic Integer Programming (SIP)** | [11] | -Optimize resource provisioning operations to reduce the total cost |
|  | **Transverse Matrix Approach** | [54] | -To minimize the data transfer time consumption between VMs (the application) and related data (the data storage). |
|  | **Convex Optimization Theory** | [62] | -To maximize the allocation of resources and minimize the transmission traffic of data center in |
| Approximation Algorithms | **O(log n)-approximation algorithm** | [25] | -To minimize the total connection costs by minimizing the length of networks connecting all physical host machines |
|  | **3-approximation algorithm** | [41] | -Obtain maximum access latency between the DNs and VMs |
|  | **Rounding Algorithm** | [14] | -Maximizing the benefit from the overall traffic sent by virtual machines to the root. |
|  | **Clustering Algorithm** | [73] | -To minimize the traffic between racks and also to decrease overall communication delay by minimizing the average communication path length, |

Table III: Classification of existing solutions


\end{*}

\bibliography{MyBib.bib}{}
\bibliographystyle{plain}

\end{document}